\documentclass[a4paper,11pt]{article}

\usepackage{pos}

\newacronym{ape}{APE}{Array Processor Experiment}
\newacronym{apbc}{aPBC}{anti-periodic boundary conditions}
\newacronym{bc}{BC}{boundary condition}
\newacronym{cg}{CG}{Clebsch-Gordan}
\newacronym{cls}{CLS}{Coordinated Lattice Simulations}
\newacronym{cm}{CM}{centre-of-mass}
\newacronym{chpt}{ChPT}{chiral perturbation theory}
\newacronym{dcsb}{DCSB}{dynamical chiral symmetry breaking}
\newacronym{ddhmc}{DD-HMC}{domain decomposition HMC}
\newacronym{ddvcs}{DDVCS}{double deeply virtual Compton scattering}
\newacronym{dis}{DIS}{deep inelastic scattering}
\newacronym{dr}{DR}{dimensional regularisation}
\newacronym{ds}{DS}{Dyson-Schwinger}
\newacronym{dvcs}{DVCS}{deeply virtual Compton scattering}
\newacronym{dvmp}{DVMP}{deeply virtual meson production}
\newacronym{eic}{EIC}{Electron-Ion Collider}
\newacronym{eicc}{EicC}{Electron-ion collider in China}
\newacronym{erbl}{ERBL}{Efremov-Radyushkin-Brodsky-Lepage}
\newacronym{fse}{FSE}{finite-size effect}
\newacronym{gcr}{GCR}{generalized conjugate residual}
\newacronym{gevp}{GEVP}{generalized eigenvalue problem}
\newacronym{hep}{HEP}{high-energy physics}
\newacronym{hmc}{HMC}{hybrid Monte Carlo}
\newacronym{ib}{IB}{isospin breaking}
\newacronym{ir}{IR}{infrared}
\newacronym{irrep}{irrep}{irreducible representation}
\newacronym{ldme}{LDME}{long-distance matrix element}
\newacronym{lhc}{LHC}{Large Hadron Collider}
\newacronym{lhs}{lhs}{left-hand-side}
\newacronym{lo}{LO}{leading order}
\newacronym{lqcd}{LQCD}{lattice quantum chromodynamics}
\newacronym{mphmc}{MP-HMC}{mass preconditioned HMC}
\newacronym{nlo}{NLO}{next-to-leading order}
\newacronym{nnlo}{NNLO}{next-to-next-to-leading order}
\newacronym{n3lo}{N3LO}{next-to-next-to-next-to-leading order}
\newacronym{nrqcd}{NRQCD}{non-relativistic QCD}
\newacronym{ope}{OPE}{operator product expansion}
\newacronym{ozi}{OZI}{Okubo–Zweig–Iizuka}
\newacronym{pbc}{PBC}{periodic boundary condition}
\newacronym{ptbc}{PTBC}{partially twisted boundary conditions}
\newacronym{qcd}{QCD}{quantum chromodynamics}
\newacronym{qed}{QED}{quantum electrodynamics}
\newacronym{qft}{QFT}{quantum field theory}
\newacronym{rgi}{RGI}{renormalization group invariant}
\newacronym{rhs}{rhs}{right-hand side}
\newacronym{sap}{SAP}{Schwarz alternating procedure}
\newacronym{sdf}{SDF}{short distance factorization}
\newacronym{slac}{SLAC}{Stanford Linear Accelerator Center}
\newacronym{svd}{SVD}{singular value decomposition}
\newacronym{tbc}{TBC}{twisted boundary condition}
\newacronym{tcs}{TCS}{timelike Compton scattering}
\newacronym{uv}{UV}{ultraviolet}
\newacronym{varpro}{VP}{variable projection}

\newacronym{gda}{GDA}{generalized distribution amplitude}
\newacronym{gpd}{GPD}{generalized parton distribution}
\newacronym{lcda}{LCDA}{light-cone DA}
\newacronym{da}{DA}{distribution amplitude}
\newacronym{pda}{pDA}{pseudo-DA}
\newacronym{pdf}{PDF}{parton distribution function}
\newacronym{ppdf}{pPDF}{pseudo-PDF}
\newacronym{itpdf}{ITPDF}{Ioffe-time PDF}
\newacronym{itd}{ITD}{Ioffe-time DA}
\newacronym{pitd}{pITD}{Ioffe-time pseudo-DA}
\newacronym{rpitd}{rpITD}{reduced Ioffe-time pseudo-DA}
\newacronym{lamet}{LaMET}{Large-momentum effecive theory}

\AtBeginDocument{\RenewCommandCopy\qty\SI}
\DeclareUnicodeCharacter{0302}{^}

\newcommand{\iu}{\text{i}}

\newcommand{\msbar}{\ensuremath{\overline{\text{MS}}}}

\title{The distribution amplitude of the $\Petac$-meson}
\ShortTitle{The distribution amplitude of the $\Petac$-meson}

\author[a]{B. Blossier}
\author[b]{M. Mangin-Brinet}
\author[c]{J.M. Morgado Ch\'{a}vez}
\author*[a]{T. San Jos\'{e}}

\affiliation[a]{Laboratoire de Physique des 2 Infinis Irène Joliot-Curie, CNRS/IN2P3,\\
Université Paris-Saclay, 91405 Orsay Cedex, France}

\affiliation[b]{Laboratoire de Physique Subatomique et de Cosmologie, CNRS/IN2P3\\
38026 Grenoble Cedex, France}

\affiliation[c]{D\'{e}partement de Physique Nucl\'{e}aire, Irfu/CEA-Saclay,\\
91191 Gif-sur-Yvette Cedex, France}

\emailAdd{blossier@ijclab.in2p3.fr}
\emailAdd{mariane@lpsc.in2p3.fr}
\emailAdd{jose-manuel.morgadochavez@cea.fr}
\emailAdd{san-jose-perez@ijclab.in2p3.fr}

\abstract{In this proceeding we determine the distribution amplitude of the $\Petac$-meson from first principles. This quantity appears as a consequence of factorization theorems, and it is necessary to compute the amplitude of multiple exclusive processes. Since it is defined along a light-cone, its calculation via lattice QCD was impossible until recently, when a generalization to Euclidean metric was proposed, and a connection to the physical limit was established. We briefly explain the method of short distance factorization, which allows us to compute the distribution amplitude, and our lattice setup. After summarizing the steps for the continuum and chiral extrapolation, we present our results and compare them to two alternative determinations, one using non-relativistic QCD and another solving the Dyson-Schwinger equations; we find a large discrepancy with the former.}

\FullConference{31st International Workshop on Deep Inelastic Scattering (DIS2024)\\
8–12 April 2024\\
Grenoble, France\\}

\begin{document}
\maketitle

\section{Introduction}
\label{sec:introduction}

Factorization theorems are a fundamental tool to study the internal structure of hadrons. They separate the amplitude of many scattering processes into two pieces: The interaction between the scattering probe and the target's quarks and gluons, which is calculable in perturbation theory; and the internal structure of the various hadrons, which requires non-perturbative methods. In particular, distribution amplitudes of heavy mesons describe the momentum distribution of the quarks along the longitudinal axis, and they are necessary to compute the amplitude of decays, deeply virtual meson production, and other exclusive processes (see \cite{Diehl:2003ny} for a review). For a pseudoscalar charmonium state like $\Petac$, its \gls{da} is defined by a bi-local matrix element \cite{Diehl:2003ny},
\begin{equation}
    \label{eq:definition-da}
    \phi(x) = \int \dfrac{\dd{z^-}}{2\pi} e^{-\iu (x-1/2) p^+ z^-} \eval{\mel*{\Petac(p)}{\APcharm(-z/2)\gamma^+\gamma_5 W(-z/2,z/2)\Pcharm(z/2)}{0}}_{z^+=z^\bot=0},
\end{equation}
where the Wilson line $W(a,b)$ assures gauge invariance, the meson moves along the axis $p^+$, the quark's momentum fraction in the longitudinal direction is $x=q^+/p^+$, and the quarks are separated along the direction $z^-$. Given its non-perturbative nature, the idea of computing this quantity \textit{ab initio} using lattice QCD arises. However, the condition $z^2=0$ prohibits the direct calculation of \cref{eq:definition-da} in Euclidean space. This was until 2013, when Ji introduced a generalization to space-like separations \cite{Ji:2013dva} known as quasi-distributions, where one could use $z=(0,0,z_3,0)$ and $p=(0,0,p_3,E)$ in Euclidean space. The new function tends towards the \gls{lcda} when taking the limit $p_3\to\infty$ appropriately. However, as it was pointed out in \cite{Radyushkin:2017cyf}, lattice simulations must include data at $p_3 \geq \qty{3}{\giga\eV}$ to take the infinite limit safely, a condition our ensembles cannot fulfill without introducing sizeable lattice artifacts. Instead, in our project we consider the equivalent approach outlined in \cite{Radyushkin:2017cyf}, known as pseudo-distributions, which converge to the physical quantity in the limit $z_3\to0$. The factor $1/z_3^2$ plays an analogous role to a renormalization parameter $\mu^2$.

In the remainder of this proceeding, we summarize how to extract the \glsxtrlong{pda} from the lattice, take the continuum limit, and match to the \gls{lcda}. Besides, we compare our results with those from \gls{ds} \cite{Ding:2015rkn} and \gls{nrqcd} \cite{Chung:2019ota} determinations. For a complete report on this work, we direct the avid reader to \cite{Blossier:2024wyx}.

\section{Methodology}
\label{sec:pseudo-da-extraction}

\begin{figure}
    \centering
    \begin{subfigure}[t]{0.49\textwidth}
        \centering
        \includegraphics{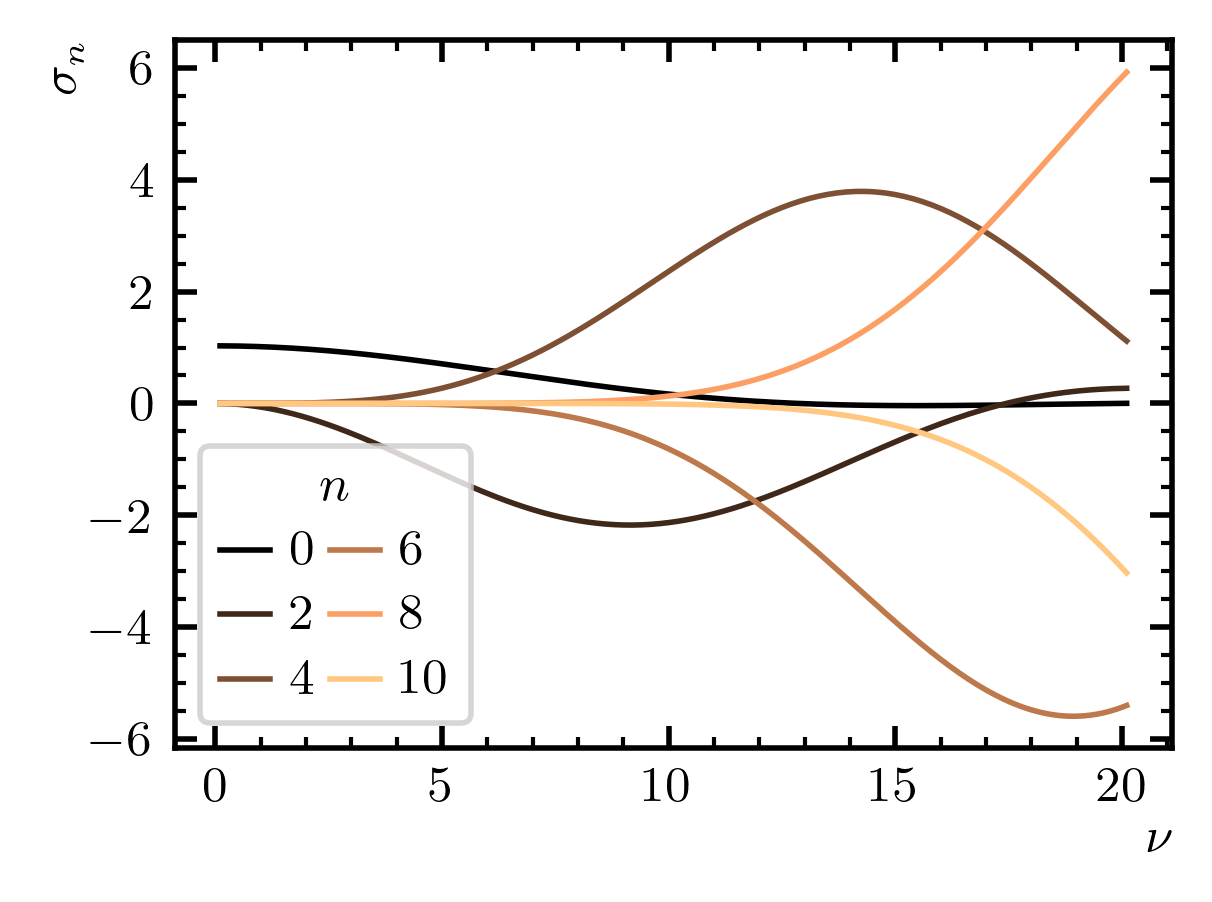}
    \end{subfigure}
    \begin{subfigure}[t]{0.49\textwidth}
        \centering
        \includegraphics{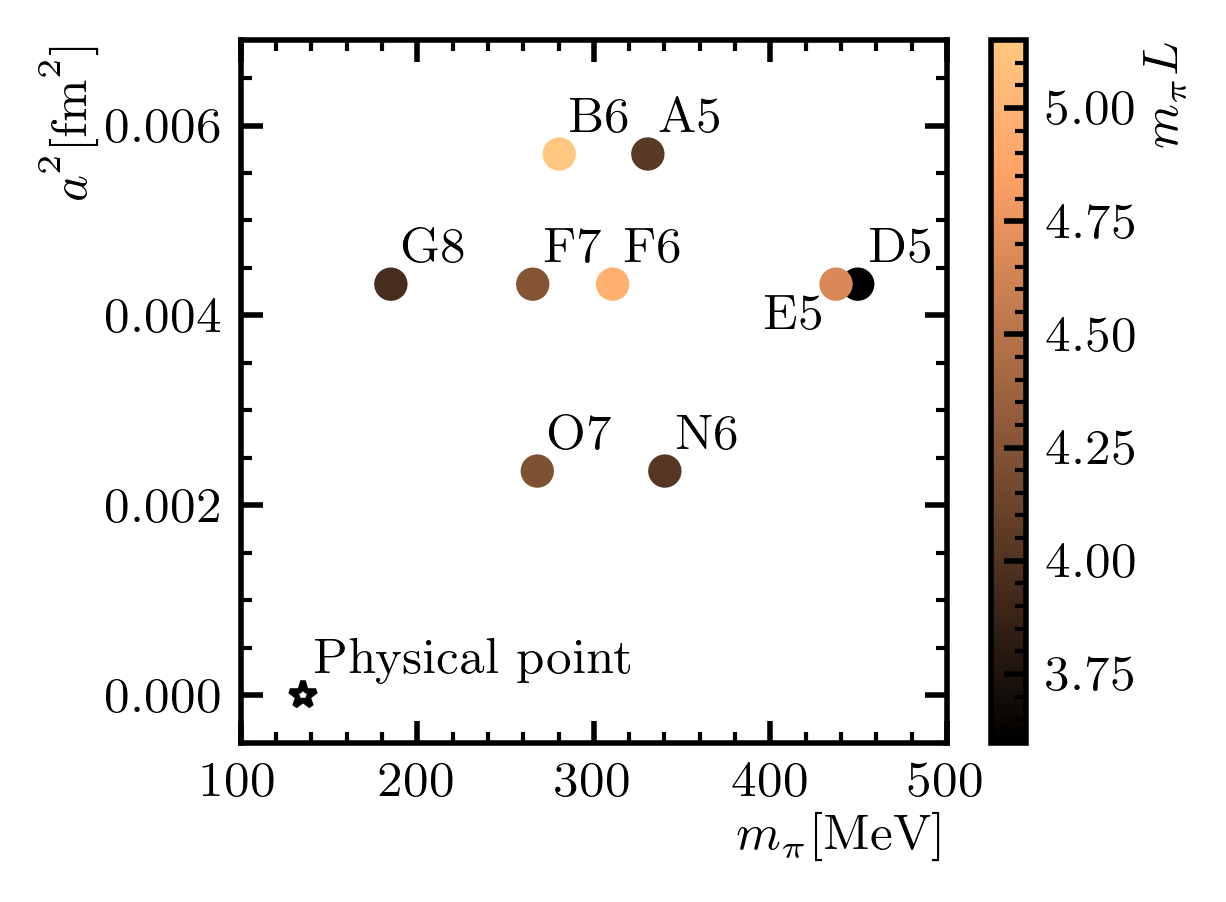}
    \end{subfigure}
    \caption{LEFT: The function $\sigma_n$ vs Ioffe time. We use $\lambda=2.7$ and $z_3 = 5 \times \qty{0.0658}{\femto\meter}$ as representative values. RIGHT: Landscape of \glsxtrshort{cls} ensembles used in this project.}
    \label{fig:sigma_n}
\end{figure}

The first step in Euclidean space is to compute the matrix element \cite{Radyushkin:2016hsy}
\begin{equation}
    \label{eq:mel-lattice}
    M^{\alpha}(p,z) = e^{-\iu pz/2} \mel*{\Petac(p)}{\APcharm(0)\gamma^{\alpha}\gamma_5 W(0,z)\Pcharm(z)}{0}
\end{equation}
where $z^2>0$ and $\nu=pz=p_3z_3$ is the Ioffe time. We consider the asymmetric configuration $(0,z)$ to make use of all lattice sites, and we recover the symmetric structure of \cref{eq:definition-da} using translation invariance. Next, we can separate some higher-twist contamination using the Lorentz decomposition \cite{Radyushkin:2016hsy}
\begin{equation}
    M^{\alpha}(p,z) = 2p^{\alpha}\mathcal{M}(\nu,z^2)+z^{\alpha}\mathcal{M}^{\prime}(\nu,z^2)
\end{equation}
where $\mathcal{M}$ carries the leading twist as well as some high twist component proportional to $\order{z^2\Lambda^2_{\text{QCD}}}$, and $\mathcal{M}^{\prime}$ is a purely higher-twist effect. To isolate $\mathcal{M}$, we select the component $\alpha=4$, such that $M^4(p,z)=2E\mathcal{M}(\nu,z^2)$. The matrix element $\mathcal{M}(\nu,z^2)$ is multiplicatively renormalizable at all orders in perturbation theory \cite{Ishikawa:2017faj}, and the normalization constant is a function of $z_3$ alone. Then, we form a ratio that retains the physical evolution of the \gls{lcda} while reducing the contamination from the lattice artifacts and higher twist \cite{Radyushkin:2016hsy,Orginos:2017kos,Karpie:2018zaz},
\begin{equation}
    \label{eq:rgi-ratio}
    \tilde{\phi}(\nu,z^2) + \text{h.t.} = \dfrac{\mathcal{M}(\nu,z^2) \eval*{\mathcal{M}(0,0)}_{p=z=0}}{\eval*{\mathcal{M}(0,z^2)}_{p=0} \eval*{\mathcal{M}(0,0)}_{z=0}}.
\end{equation}
\Cref{eq:rgi-ratio} is the \glsentrylong{rpitd}, which does not require any additional renormalization, and so it has a well defined continuum limit. This is the actual quantity we extract from the lattice data. \Cref{eq:rgi-ratio} is related to the \gls{da} on the light-cone via a matching kernel derived in 
\cite{Radyushkin:2017lvu,Radyushkin:2019owq} at \glsxtrshort{nlo} in perturbation theory,
\begin{equation}
    \label{eq:pseudo-to-light-cone}
    \tilde{\phi}(\nu,z^2) = \int_0^1 \dd{w} C(w,\nu,z\mu) \int_0^1 \dd{x} \cos(wx\nu-w\nu/2) \phi(x,\mu),
\end{equation}
where we choose the $\msbar$ renormalization scheme at a scale $\mu=\qty{3}{\giga\eV}$. \Cref{eq:pseudo-to-light-cone} is an inverse problem, because we attempt to reconstruct the \glsxtrlong{rhs} with only a limited data set. To solve it, we model the \gls{lcda} taking inspiration from its conformal expansion \cite{Diehl:2003ny},
\begin{equation}
    \label{eq:light-cone-da-model}
    \phi(x,\mu)
    = (1-x)^{\lambda-1/2}x^{\lambda-1/2} \sum_{n=0}^{\infty} d_{2n}^{(\lambda)} \tilde{G}_{2n}^{(\lambda)}(x),
    \qquad\qquad
    d_0^{(\lambda)} = \dfrac{4^{\lambda}}{B\left(1/2, \lambda+1/2\right)}.
\end{equation}
The coefficients $d_{2n}^{(\lambda)}$ and $\lambda$ are fitted with the lattice data, $B$ is a beta function, and $\tilde{G}_{2n}^{(\lambda)}(x)$ are shifted Gegenbauer polynomials, defined in the domain $x\in[0,1]$. The assymptotic value, $\phi(x,\mu\to\infty)=6x(1-x)$ is recovered from \cref{eq:light-cone-da-model} setting $\lambda=3/2$. A similar approach for \glspl{pdf} can be seen in \cite{Karpie:2021pap}. Then, we can replace \cref{eq:light-cone-da-model} in \cref{eq:pseudo-to-light-cone} and compute the integrals using properties of the Gegenbauer polynomials, such that
\begin{equation}
    \label{eq:matching-series}
    \tilde{\phi}(\nu,z^2)
    =
    \sum_{n=0}^{\infty}\tilde{d}_{2n}^{(\lambda)}\sigma_{2n}^{(\lambda)}(\nu,z\mu),
    \qquad\qquad
    \tilde{d}_n^{(\lambda)} = \dfrac{d_n^{(\lambda)}}{4^{\lambda}}.
\end{equation}
The new set of functions $\sigma_{2n}^{(\lambda)}$ are plotted in \cref{fig:sigma_n} for representative values of our simulations, and their full expressions are reported in \cite{Blossier:2024wyx}.

\section{Lattice calculation}
\label{sec:lattice-calculation}

\begin{figure}[htbp]
    \centering
    \includegraphics{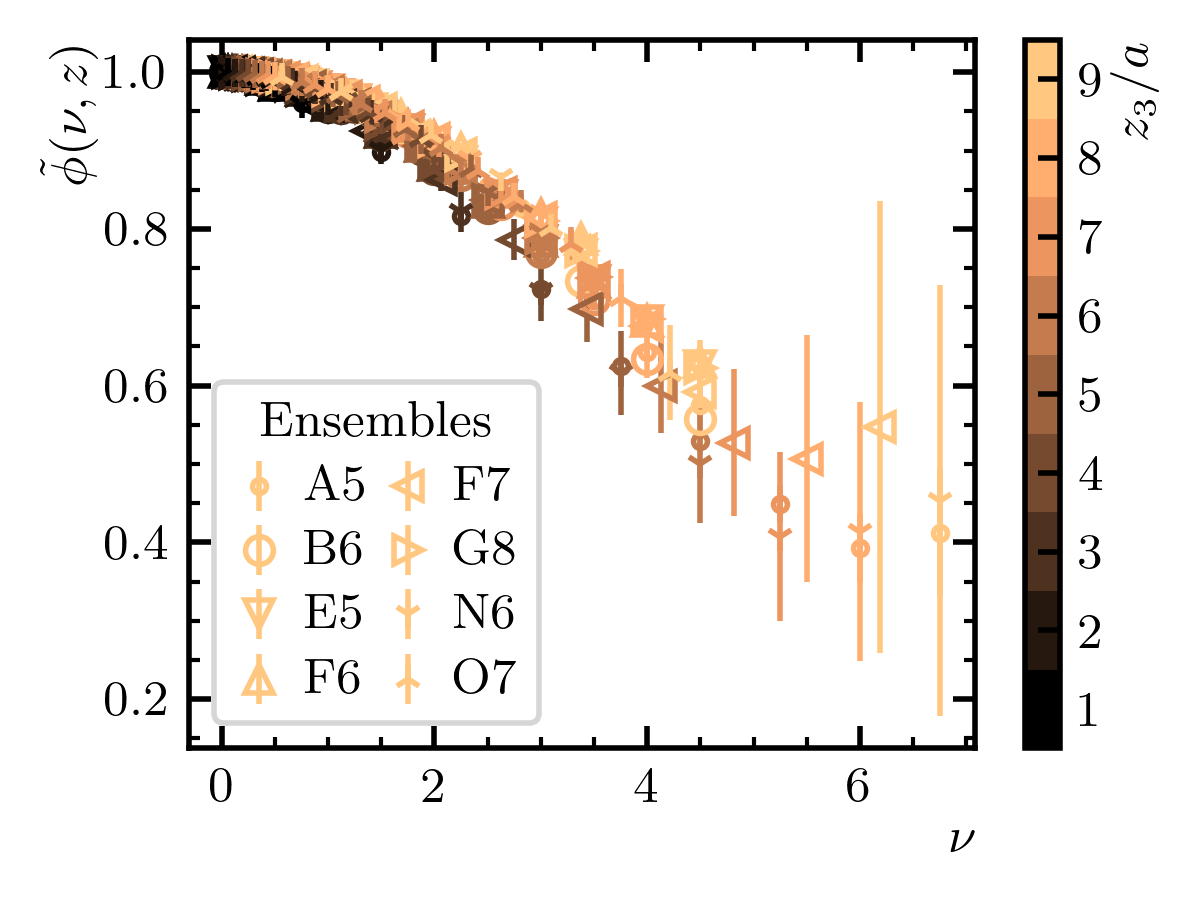}
    \caption{The \glsxtrlong{rpitd} as a function of Ioffe time on all our ensembles, which are shown with different markers. Different Wilson lines are given by the color gradient.}
    \label{fig:rpitd}
\end{figure}

\begin{figure}
    \centering
    \includegraphics{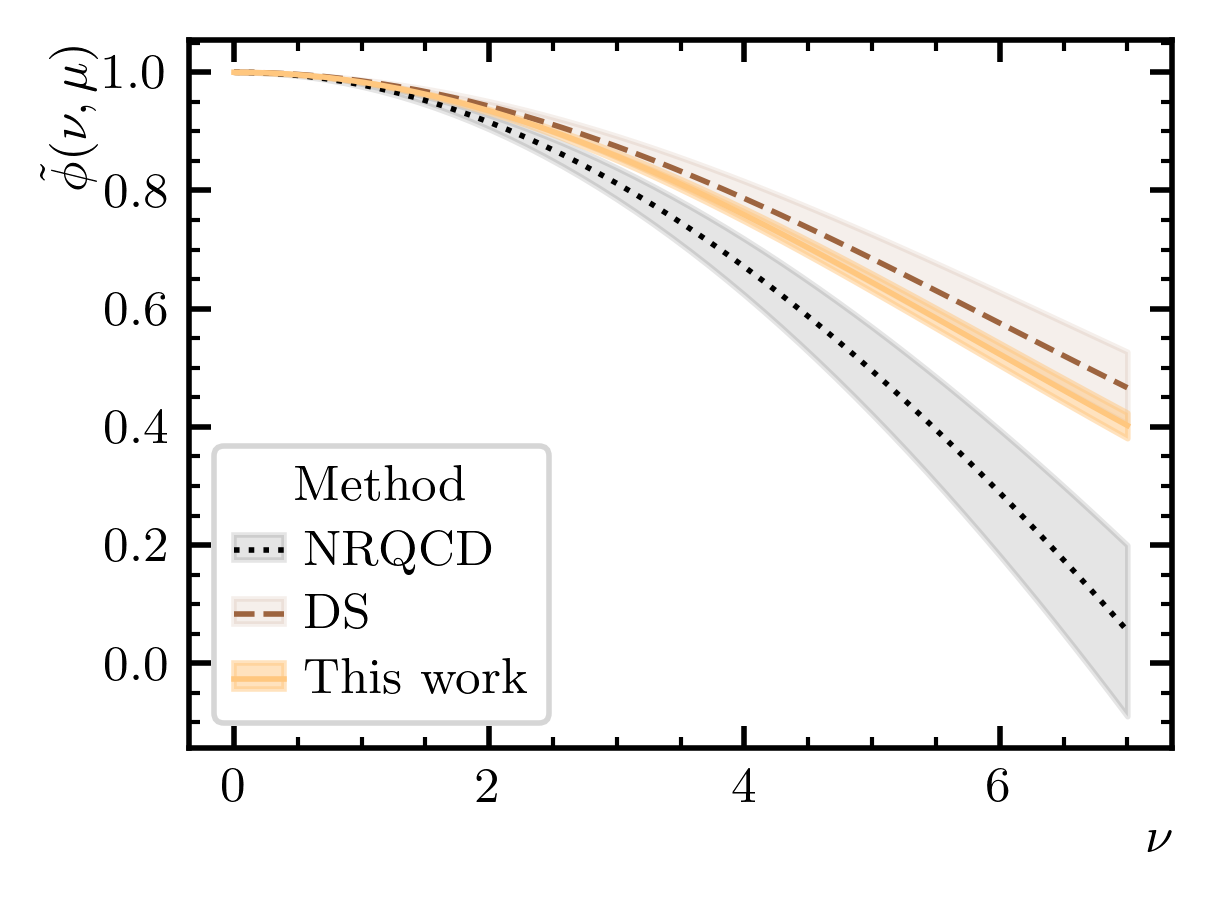}
    \caption{Comparison in Ioffe-time space between the \gls{da} as determined in this work, by \gls{nrqcd} \cite{Chung:2019ota} and by \gls{ds} \cite{Ding:2015rkn}. The bands reflect the uncertainties reported on each work, ours indicate the total error.}
    \label{fig:comparisons}
\end{figure}

We employ the $N_f=2$ \glsxtrshort{cls} ensembles collected in \cref{fig:sigma_n}. For details on the gauge simulations and the scale setting, see \cite{Fritzsch:2012wq}. We employ quark all-to-all propagators with wall sources diluted in spin, and we solve the Dirac equation with deflated \glsxtrshort{sap}-\glsxtrshort{gcr}, and the contractions are performed with a costum version of the \gls{ddhmc} algorithm. The momentum $p$ is set employing \glsxtrlong{ptbc}, and to create a realistic meson interpolator we solve a 4x4 \glsxtrlong{gevp} with four levels of Gaussian smearing. Out of the two Wick contractions, we only compute the quark-connected diagram, as we expect the disconnected contribution to have a strong \glsxtrshort{ozi} suppression. Upon averaging over the spatial lattice points, the double ratio in \cref{eq:rgi-ratio} shows a mild time dependence, and we select $\tilde{\phi}$ picking a particular time slice from a plateau range. Selecting the same time slice for all $(\nu,z^2)$ on a given ensemble, we can keep correlations intact throughout our analysis, and varying the selected time slice we explore the systematic uncertainty \cite{Blossier:2024wyx}. \Cref{fig:rpitd} shows \cref{eq:rgi-ratio} on all our ensembles. We observe that all data points, which we denote by $\tilde{\phi}_e(\nu,z^2)$, fall in a nearly universal line, but several corrections remain to be applied if one is to extract \cref{eq:definition-da}: First, we need to model the quark-mass dependence to give a prediction at the physical value; second, we need to take the continuum limit removing the $\order{a}$ lattice artifacts; third, we have to model the remaining higher-twist contamination.
We combine these steps in a single fit, where we minimize a $\chi^2$ employing a custom implementation of the \glsxtrlong{varpro} algorithm, and we fit our data to the model \cite{Blossier:2024wyx}
\begin{equation}
  \label{eq:continuum-extrapolation-model}
  \begin{aligned}
    \tilde{\phi}_e(\nu,z^2)
    &=
    \tilde{\phi}(\nu,z^2) + \dfrac{a}{\abs{z}} A_1(\nu) + a\Lambda B_1(\nu) + z^2\Lambda^2 C_1(\nu)
    \\
    &+\dfrac{a}{\abs{z}}
      \Big(
        \Lambda^{-1} \left(m_{\Petac} - m_{\Petac,\text{phy}}\right) D_1(\nu)
        + \Lambda^{-2} \left(m_{\Ppi}^2 - m_{\Ppi,\text{phy}}^2\right) E_1(\nu)
      \Big),
  \end{aligned}
\end{equation}
which gathers all the aforementioned requirements. See \cite{Karpie:2021pap} for a study of \glspl{pdf} using a similar approach.
We use $\Lambda\equiv\Lambda^{(2)}_{\text{QCD}}=\qty{330}{\mega\eV}$ \cite{FlavourLatticeAveragingGroupFLAG:2021npn} to render all terms dimensionless. The auxiliary functions $A_1$, $B_1$, $C_1$, etc., have a similar behavior in $\nu$ to that of $\tilde{\phi}(\nu,z^2)$, and they are defined in very similar fashion to \cref{eq:light-cone-da-model}. Since $0 < \nu < 7$, our fit function is only sensitive to the first coefficient in \cref{eq:light-cone-da-model}, and we limit ourselves to determine the value of $\lambda$. At the minimum $\chi^2/\text{dof}=368/467=0.79$, we obtain $\lambda=\num{2.73(18)}$, where the uncertainty includes both statistics and systematics. This is our main result, and we stress that knowing the first coefficient in \cref{eq:light-cone-da-model} is sufficient to describe the \gls{da} in our range of Ioffe times based on the good fit quality.
The \gls{da} is analytic in Ioffe-time space, and therefore we Fourier-transform \cref{eq:light-cone-da-model} to compare it to two other determinations using \gls{nrqcd} \cite{Chung:2019ota} and \gls{ds} equations \cite{Ding:2015rkn}. The various results are gathered in \cref{fig:comparisons}, and we observe good agreement between our result and \cite{Ding:2015rkn}, but there are large discrepancies with the non-relativistic calculation.

\section{Conclusions and Outlook}
\label{sec:conclusions}

We present the first lattice calculation of the $\Petac$-meson \gls{da}. We give the \gls{da} in closed form using \cref{eq:light-cone-da-model} truncating $n>0$ and using $\lambda=2.73(18)$. This parameterization is sufficient to fit the lattice data, which is generated following the pseudo-distribution approach. Our results are obtained in the continuum limit at the physical quark masses (with exact isospin symmetry) and at leading twist. Our determination is in strong tension with results from \gls{nrqcd}. In the future, we plan to expand our analysis to other heavy states as well as to upgrade our simulations to $N_f=2+1+1$ flavors in the sea.

\acknowledgments

The work by J.M. Morgado Ch\'{a}vez has been supported by P2IO LabEx (ANR-10-LABX-0038) in the framework of Investissements d’Avenir (ANR-11-IDEX-0003-01). The work by T. San Jos\'{e} is supported by Agence Nationale de la Recherche under the contract ANR-17-CE31-0019. This project was granted access to the HPC resources of TGCC (2021-A0100502271, 2022-A0120502271 and 2023-A0140502271) by GENCI. The authors thank Michael Fucilla, Cédric Mezrag, Lech Szymanowski, and Samuel Wallon for valuable discussions.

\bibliographystyle{JHEP}
\bibliography{bib}

\end{document}